\documentclass[journal, 10pt]{IEEEtran}
\usepackage{graphicx,subfig}
\usepackage[cmex10]{amsmath}
\usepackage{algorithmic ,algorithm}
\usepackage{array}
\usepackage{mdwmath}
\usepackage{mdwtab}
\usepackage{multirow}
\usepackage{color}

\begin{document}
\title{Virtual Machine Migration Enabled Cloud Resource Management: A Challenging Task }
\author{Misbah Liaqat$^a$, Shalini Ninoriya$^b$, Junaid Shuja$^a$, Raja Wasim Ahmad$^a$, Abdullah Gani$^a$ 
\\$^a$ Faculty of Computer Science and Information Technology, FSKTM, University of Malaya
\\ $^b$ Guru Ramdas Khalsa Institute of Science and Technology Jabalpur\\Email: mibahaseeb@siswa.um.edu.my, shalini\_2\_8@yahoo.co.in, junaidshuja@siswa.um.edu.my, wasimraja@siswa.um.edu.my, abdullah@um.edu.my} 

\maketitle

\begin{abstract}
Virtualization technology reduces cloud operational cost by increasing cloud resource utilization level. The incorporation of virtualization within cloud data centers can severely degrade cloud performance if not properly managed. Virtual machine (VM) migration is a method that assists cloud service providers to efficiently manage cloud resources while eliminating the need of human supervision. VM migration methodology migrates current-hosted workload from one server to another by either employing live or non-live migration pattern. In comparison to non-live migration, live migration does not suspend application services prior to VM migration process. VM migration enables cloud operators to achieve various resource management goals, such as, green computing, load balancing, fault management, and real time server maintenance. In this paper, we have thoroughly surveyed VM migration methods and applications. We have briefly discussed VM migration applications. Some open research issues have been highlighted to represent future challenges in this domain. A queue based migration model has been proposed and discussed to efficiently migrate VM memory pages. 
\end{abstract}

\begin{IEEEkeywords}
Virtualization, Live migration, Data centers, energy efficiency.
\end{IEEEkeywords}

\section{Introduction}\label{sec:intro}
Information Technology (IT) industry has evolved from its birth in the last century into one of the most prominent industry in today's world. Along with its rapid growth, IT is changing daily lifestyles and is becomimg a technology enabler for many veteran industries and businesses~\cite{Zeadally2011}. Cloud computing technologies have emerged as a backbone of all IT services. Cloud computing traditionally used to provide infrastructure, platform, software, and data as a service. Due to the immense popularity of cloud computing technologies, it is envisioned now that the cloud will provide on-demand Everything as a service architecture for users worldwide in the future~\cite{Banerjee2011a}. Cloud computing has also been termed as the fifth utility for human beings~\cite{Buyya2009a}. The cloud facilities utilize a large number of computing resources networked together in the form of warehouse size data centers. Clients utilize these computing resources on-demand through managed cloud applications with pas-as-you-go model. The on-demand utility based cloud model leads to lower operating costs, small investments, high scalability, and multi-tenancy characteristics that are ideal for small industries and businesses~\cite{Zhang2010}. The recent on-demand availability of computing resources in the form of cloud has led to a major technological and economical revolution in the IT industry.  

Energy efficiency is a global challenge for today's world. Conventional energy resources are being consumed rapidly while call for alternative energy resources is growing from different industries. IT technologies, including cloud computing, have become a major consumer of energy due to rapid growth and integration with other industries~\cite{Shuja2012}. Therefore, many IT technologies have focused on saving energy by developing energy efficient protocols, architectures, and mechanisms at cloud back end data centers~\cite{Zeadally2011,Ahmed2015a}. Although cloud computing is termed as inherently energy efficient platform due scalable nature of its resources and multi-tenant capability, densely populated data centers are usually over provisioned and consume large amounts of energy in a competitive race between the service providers to ensure 24/7 availability of services to the clients~\cite{Kliazovich2010}. Moreover, the cloud technologies and back end data center utilizing grid electricity are responsible for fair amount of CO2 emission~\cite{Gmach2010}.      \begin{figure*}
\centering
\includegraphics[width=12cm]{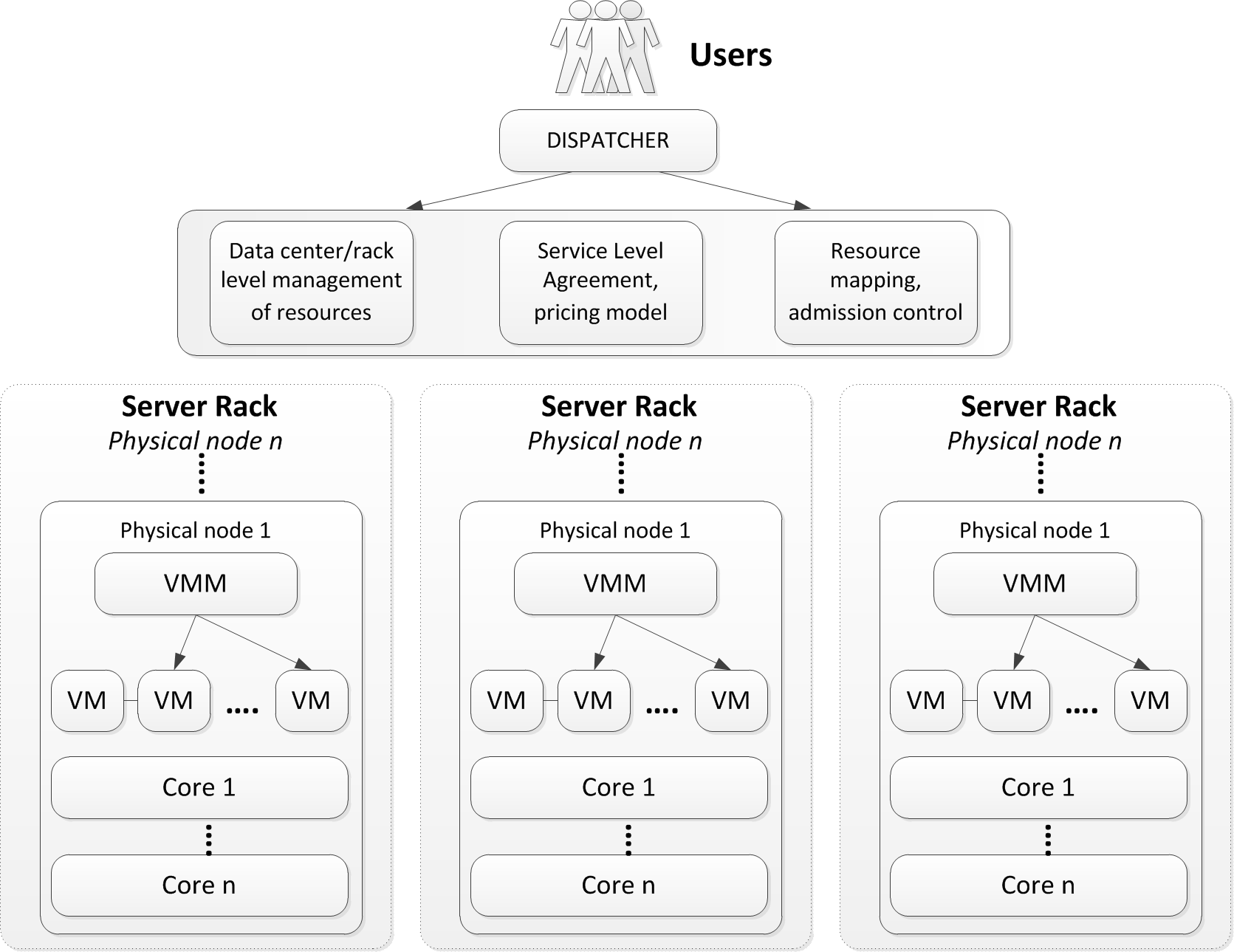}
\caption {Architecture of a virtual data center}
\label{fig:vmm}
\end{figure*} 
This paper comprehensively discusses virtual data center architecture, state-of-the-art VM migration techniques, VM migration applications, Issues and recommendations. The organization of the paper is as follows. Section~\ref{sec:architecture} comprehensively discusses virtual data center architecture. Section~\ref{sec:lit} discusses state-of-the-art literature regarding VM migration schemes and models. Section~\ref{sec:apps} discusses VM migration applications. Moreover, Section~\ref{sec:issues} discusses open research issues in VM migration domain. Section~\ref{sec:model} proposes Queue based VM migration model. Section~\ref{sec:conc} concludes the whole paper and describes the future work.  

\section{Virtualized data center architecture}\label{sec:architecture}
Virtualization was devised as a resource management and optimization technique for mainframes having scaleless computing capabilities. Virtualization in mainframes results in efficient management of coarse-grained resources with limited overhead. However, virtualization techniques have been able to make their way to multi-core~\cite{Petrides2012} and commodity server designs~\cite{Egi2010}. The multi-core processor, blade server, and System on Chip (SoC) designs also provide virtualization techniques opportunity for resource consolidation and optimization where fine-grained resources are assembled to provide a virtual scalable platform for cloud applications. Virtualization techniques benefits data centers in many ways, such as:

\begin{itemize}
  \item \textbf{Scalability:} cloud users and applications view heterogeneous hardware and software resources as a single scalable platform. Virtual devices are scalable in terms of hardware resources. Server, memory, and I/O power can be added to a virtual machine when its resource utilization nears 100\%.
  \item \textbf{Consolidation and utilization:} virtual resources can be easily consolidated over few physical resources that results in higher resource utilization levels and energy efficiency\cite{Younge2011}.
  \item \textbf{Isolation:} performance and faults are isolated between applications of the same resource~\cite{Uhlig2005}. 
  \item \textbf{Manageability:} virtualization offers variety of resource management options such as VM creation, deletion, and migration.
	\item \textbf{Robustness:} virtualization leads to system robustness as clients spread across multiple VMs. 
\end{itemize}  

Due to the aforementioned benefits, virtualization is globally adopted in cloud data center environments~\cite{Goiri2012,Ahmed2015}. In a virtualized data center architecture, each cloud client (application or user) is assigned a chunk of data center resources. The data center resources close to the hardware platform can be categorized into physical resource set and virtual resource set~\cite{Lenk2009}. The virtual resource set works as a management platform over the physical resource set to provide the illusion of a single scalable platform to all cloud clients. A hypervisor or Virtual Machine Monitor (VMM) is hardware independent technology that manages virtual machines over heterogeneous hardware platforms. The hypervisor is a set of computer hardware, firmware, and software that lies between the hardware and the Operating System (OS). The hypervisor has the ability to initiate one or more than one OSes over a single hardware resource set~\cite{Younge2011}. Inside a virtualized data center, clients reside over a pool of virtual resource sets. When a new client request arrives at the cloud data center, it is forwarded by the dispatcher to corresponding VMM. The dispatcher requests physical resources according to the client SLA, pricing model, and application QoS requirements~\cite{Buyya2011,shiraz2015energy}. However, escalating energy costs of cloud resources has added another dimension to admission control of client requests: resource and workload consolidation for energy efficiency. The resource mapping can be based on virtual machine placement~\cite{Urgaonkar2010}, thermal profile~\cite{Pakbaznia2010}, or network characteristics of the data center~\cite{Kliazovich2010a}. Moreover, virtual machine placement can be optimized  as energy aware~\cite{Dupont2012,Shuja2014,Shuja2014a}, traffic patterns aware~\cite{Meng2010}, network congestion aware~\cite{Piao2010}, or can be multi-objective~\cite{Xu2010}. Figure~\ref{fig:vmm} illustrates the architecture of a virtualized data center consisting of multi-core servers. 

\begin{figure}
\centering
\includegraphics[width=8cm]{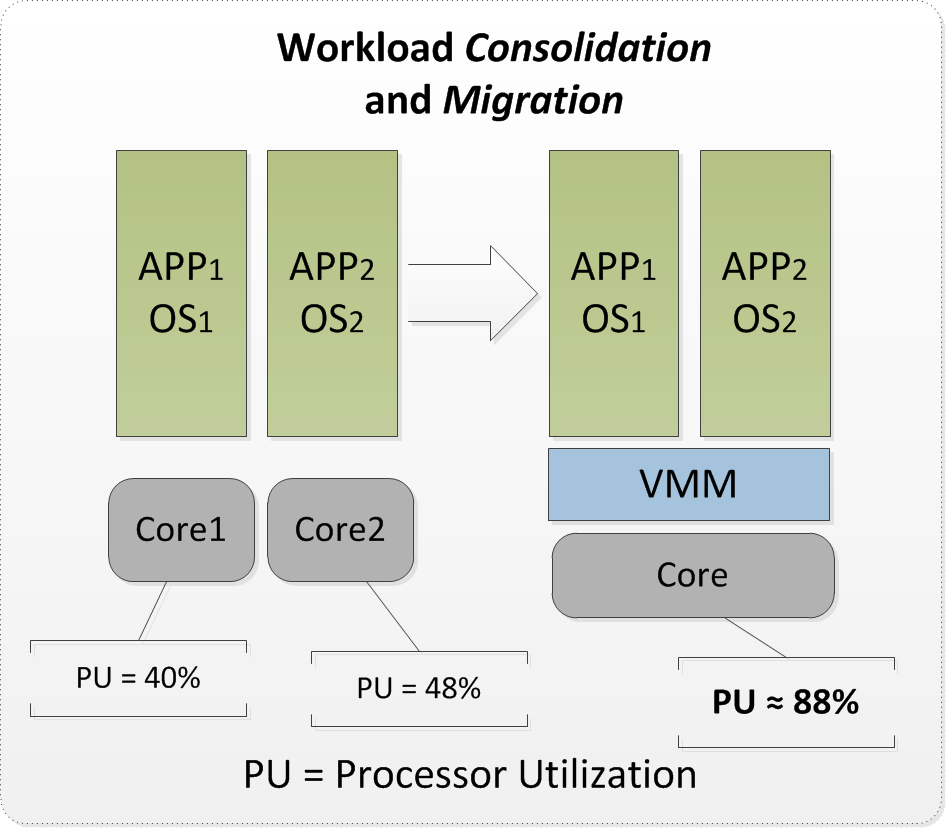}
\caption {Workload consolidation and migration}
\label{fig:mig}
\end{figure}

In a virtualized data center architecture, multiple clients often share same hardware resources with the help of virtualization techniques. Moreover, hardware resources provisioned for a data center client can be scaled dynamically according to varying workload. The resource scaling can be done with a variety of virtualization methods such as VM creation, deletion, and migration. A workload can be consolidated or migrated onto a lesser number of resources using VM migration. The resultant resource set provides energy efficiency and higher resource utilization. The CPU power along with other computing resources such as memory and I/O can be scaled gracefully with the help of virtualization technologies~\cite{Younge2011}. When a hardware resource is underutilized due to lesser client requests, it represents and opportunity for resource consolidation. The workload of underutilized hardware is transferred to another suitable hardware with the help of hypervisor. The workload consolidation and migration technique is depicted in figure~\ref{fig:mig}.

\subsection{Virtualized data center controller}\label{sec:soft}
The state of the VM that needs to be transferred to the destination includes \textbf{(a)} memory (probably RAM), \textbf{(b)} state of devices such as CPU and NIC (CPU registers, IP address migration), and \textbf{(c)} storage (HDD). Parameters to be deliberated in VM live migration are, (a) migration time, lesser the better, (b) network traffic utilization by migration, and (c) application performance degradation~\cite{Nicolae2012}. Hardware resource used by the application and for VM migration are same, so application performance degrades.
   
\section{State-of-the-art}\label{sec:lit}
Voorsluys et al.~\cite{Voorsluys2009} have used a testbed of a head node and 5 VM hosts to study the effects of VM live migrations in a data center hosting Web 2.0 applications. The VM hosts run olio\footnote{http://incubator.apache.org/olio} that defines a simple Web 2.0 application while Faban load generator\footnote{http://faban.sunsource.net} is used for workload generation. A downtime of 3 seconds was observed near the end of a 44 second migration. Although no request was dropped during the downtime, the delay does affect the service level agreement (SLA). While each VM had a 2GB of memory allocation in the experimental setup, in real data center environments the size of a VM can scale up to hundreds of GB. Therefore the effects of migrating large VM's can be more severe. Furthermore the results showed that 2 VMs migrations occurring in close time proximity lead to sever SLA violations. Hence the modeling of live migration as a queuing system comes under consideration.

Stage et al.~\cite{Stage2009} discuss the impact of VM live migration on the network resources. The proposed architecture consists of VM workload classifier, an allocation planner, a non-conformance detector, and a live migration scheduler. The VM workload classifier assigns the workload to a relevant cluster class based on the attributes of the workload. The allocation planner determines the resource bottlenecks that can occur after an allocation. The non-conformance detector classifies the bottlenecks detected on the basis of pre-defined performance thresholds. The migration requests are made by allocation planner and non-conformance detector to the migration scheduler. A migration scheduler determines the optimal schedule for the migrations, based on the knowledge of their duration, starting time and deadline. The optimal scheduler schedules the live migrations in such a way that the network is not congested by the VM live migration load. The live migrations are also fulfilled in time. The following diagram shows the difference between an uncontrolled and controlled migration scheduling algorithm. The lower timeline depicts a controlled migration in which three migration requests are executed as compared to two requests in an uncontrolled environment.

Beloglazov et al.~\cite{Beloglazov2010} have proposed that live migration of VMs can be used to concentrate the jobs on a few physical nodes so that the rest of the nodes can be put in a power saving mode. The allocation of VMs is divided into two sub-problems: (a) the admission of new requests and (b) optimization of current VM allocations. The allocation, of new requests for VMs, is done by sorting all the VMs in a Modified Best First Decreasing (MBFD) order with respect to the current utilization. The VM is then allocated to a host based on the least deterioration in the power consumption among the hosts. The current allocation of VMs is optimized by selecting the VMs to be migrated on the basis of heuristics related to utilization thresholds. If the current utilization of a host is below a threshold, then all the VMs from that host should be migrated and the host is put in the power saving mode. Again the allocation of VMs to hosts is done by MBFD algorithm. 

A similar approach achieves energy efficiency with the help of Limited Look Ahead Control (LLC)~\cite{Kusic2008}. The LLC predicts the next state of the system by a behavioral model that depends on the current state, environment input and control input. A profit maximization problem, based on the non-violation of SLA and the energy conservation, is formulated to calculate the maximum number of physical hosts that can be powered off. The optimization problem suffers the curse of dimensionality as more control options and longer look ahead horizon are considered during formulation. To avoid the curse of dimensionality, the problem is decomposed into two sub-problems with respective sub-controllers. Although this approach caters for most of the virtualized environment dynamics, such as SLA and energy efficiency, it does not consider the effects of live migration on network dynamics.
 
In SecondNet~\cite{Guo2010}, a central Virtual Data Center (VDC) manager controls all the resources and VM requests. When the VDC manager creates a VM for the VDC, it assigns the VM, a VDC ID and a VDC IP address, reserves the VM-to-VM and VM-to-core bandwidths, as mentioned in the Service Level Agreement (SLA) for the application using the VM. The inputs to the VM allocation algorithm are the \textit{m} VMs and the ${m*m}$ bandwidth matrix $R^{9}$. The output is \textit{m} physical server and the paths corresponding to the bandwidth matrix $R^{9}$. Cluster of servers are formed based on the number of hops from one cluster to another. A cluster is chosen, $C_{k}$, such that: (a) it has more ingress and egress bandwidth then that specified in $R^{9}$ and (b) the number of servers in the cluster is larger than the number of VMs i.e. \textit{m}. A bipartite graph is formed from the VMs \textit{(m)} and physical servers in the cluster $C_{k}$. Mapping from VMs \textit{(m)} are made to physical hosts in $C_{k}$ based on individual VMs memory, CPU and bandwidth requirements. A bandwidth defragmentation algorithm is also devised to reduce inter-cluster bandwidth and improve network utilization. A VM migration is scheduled if meets the following criteria \textbf{(a)} it increases the residual bandwidth of the data center and \textbf{(b)} the bandwidth requirements can be met by the cluster where VMs are reallocated. Simulations demonstrate that the system provides a guaranteed bandwidth and high network utilization. This approach does consider the residual bandwidth for VM allocation optimization, but it does not consider the bandwidth required during the process of reallocation.

A study to measure the impact of virtualization on network parameters, such as throughput, packet delay, and packet loss has been conducted by Wang et al.~\cite{Wang2010}. The study is carried out on the Amazon EC2 data center where each instance of the data center is a Xen VM. Processor utilization and TCP/UDP throughput are measured by CPUTest and TCP/UDPTest programs, respectively. The packet loss is measured by the Badabing tool~\cite{Sommers2005}. The results show an unstable TCP/UDP throughput and a very high packet delay among EC2 instances. It is concluded that these results are obtained due to virtualization and sharing of drivers among several VMs.

The above discussion concludes that although virtualization is a widely adopted power management and resource allocation technique, the impact of virtualization and VM live migrations has to be considered to reduce the network performance indicators. Furthermore, live migrations can be adopted to reduce VM-to-VM traffic and such a criterion should be built into the current VM based energy-efficiency solutions.

\section{VM Migration Applications}\label{sec:apps}
This section debates on VM migration applications and presents state-of-the-art schemes to highlight their importance. 
\subsection{VM Migration Enabled Load Balancing}\label{sec:vm}
Load balancing, a deployed function, plays its vital role in cloud and cloud data center domains for efficient resource management. Load balancing ensures even distribution of resources among a set of users in a uniform way such that underlying servers do not become overloaded and idle at any time within cloud operation time line. Overlooking load balancing establishment abruptly decreases system throughput due to overloaded servers and ultimately leads to SLA violation. It has become an integral part of all distributed internet based systems as distributed computing comes with the challenges of high resource demands that overload servers. Load balancer increases the capacity and reliability of applications by decreasing the burden on a server. To achieve the load balancing, several load balancing schemes such as Minimum Execution Time (MET), Min-Min scheduling, Cloud Analyst have been reported in literature in addition to a comprehensive study on First Come First Serve (FCFS) and Round-robin schedulers. Load balancer starts with identification of hot spot, an overloaded server, and start migrating its load on a server which has sufficient resources such that the resources are evenly distributed. However, the criterion of where, which, and how to migrate workloads from the physical servers pose challenges that cloud operator has to consider during all these decision makings  ~\cite{ahmad2015survey,ahmad2015review}. 

The VectorDot scheme as discussed in~\cite{mishra2011theory} has considered the current load on the communication paths connecting physical servers and network attached storage. Furthermore, VectorDot has addressed the overloaded servers, switches, and storage entities while meeting the desired objective function. Moreover, using constraint programming paradigm, tasks are migrated within nodes located in a cluster and has proved that consolidation overhead is indomitable while choosing a new configuration and also it is affected from the total migration time with that configuration~\cite{hermenier2009entropy}. Furthermore, employed Entropy has  significantly reduced total VM migration duration in addition to the total number of nodes acquiring low performance overhead. Consequently, the authors of~\cite{zhao2007experimental} has accurately projected the total migration cost in order to have an accurate estimation guess of migration time, so that sufficient resource can be prepared and reserved on the basis of VMs count and the performance degradation period instigated by VM migration, that is higher than actual total migration duration. Moreover, the proposed scheme has also presented the migration cost based on the migrating VM configuration and size. 

\subsection{Green Data Centers}\label{sec:green}
The digital activities of the human beings are generating a large amount of data. The data needs to be processed, stored, and transmitted to the users on daily bases. These activities lead to large amount of energy consumption and have an environmental impact in form of Greenhouse Gas (GHG) emissions. CDCs are a major consumer of electricity and a major producer of GHGs. CDCs emit GHGs indirectly during the process of electricity generation, data center equipment manufacturing, and disposal. Data centers were responsible for 2\% of global emissions in 2011. The main technique to lower the carbon emissions of CDCs is the use of renewable energy~\cite{Shuja2014}. Currently, the capital cost of renewable energy is quite high. However, it is predicted that in a few years, the capital cost of renewable energy will reduce, thereby necessitating application in many industries. Another challenge to application of renewable energy to CDCs is its unpredictable production. Therefore, hybrid power supply architectures are required with renewable energy powered CDCs. Automatic Transfer Switches (ATS) are required to switch the CDC power supply between the grid energy and the renewable energy source~\cite{ahmad2015survey,Shuja2014,Shuja2014a}.

Sustainablility and green resources in cloud data centers can be integrated by three techniques. Firstly, dynamic load balancing enabled by inter-data center VM migration can be practiced to operate the cloud servers on the available renewable energy. Renewable energy resources such as solar and wind energy can be either installed on-site or purchased from grid. Secondly, intra-data center VM migrations can be utilized to migrate data center workload between geo-dispersed cloud data centers to optimally utilize renewable energy. In this manner, when renewable energy is not available at one data center site, VM migrations can help to relocate workload elsewhere. Thirdly, server power can be capped or limited to keep a balance between renewable power generation and utilization. However, this technique will increase the wait time for clients. Green Cloud Computing improves the utilization of underlying computing resources that cloud consumer uses, including physical servers, network attached storage, cloud applications, and services while reducing energy consumption of all the aforementioned entities. 

\section{Research Issues}\label{sec:issues}
Virtualization is a technique that allows the sharing of one physical host among multiple virtual machines (VM), where each VM can serve different applications. The CPU and memory resources can be dynamically provisioned for a VM according to the current performance requirements. This makes virtualization perfectly fit for the requirements of energy-efficiency in data centers. Virtualization is the most adopted power management and resource allocation technique used in cloud computing infrastructure and data centers~\cite{Chernicoff2009}. Virtualization in network domain does not provide for energy-efficiency. In effect, network resources are burdened by the virtualization techniques. Live migration of VMs in the data center is an active area of research as data has to be transferred from one physical host to another, generating a significant amount of traffic~\cite{Voorsluys2009}. The live migration of a VM essentially requires the copying of VM memory pages from the current location to a new location across the network while the VM does not stop its services at the current location. The pages that are modified during the process of live migration are marked as dirty and have to be re-transferred after the first iteration of the copy.

In virtual data center infrastructures, the goals of energy-efficiency and resource utilization arise along with the problem of non-optimized placement of VMs on different physical hosts. The non-optimal placement of VMs results in two VMs with large mutual (VM-to-VM) traffic being placed in different network domains with multiple-hop distances. The VM-to-VM traffic consumes a significant part of available network bandwidth. Energy-efficiency, higher resource utilization and optimal VM placement can be achieved by VM live migrations. The main disadvantage of live migration is that it can consume significant network bandwidth during the process of VM memory image transfer from one physical host to another.  

\begin{itemize}
  \item Transferring large sized data over the shared network link is a big challenge, especially when several goals in terms of SLA violation avoidance, minimum end to end delay, high throughput, and high service quality has to meet. 
  \item State-of-the-art VM migration schemes suppress VM contents using de-duplication, compression, write throttling, and various innovative ways (workload enabled compression) to efficiently utilize bandwidth capacity. However, applying all theses optimizations consumes significant  amount of system resources which ultimately affects co-hosted application performance in terms of SLA violation.
  \item For effective resource utilization, VMs are packed on a few servers. However, the decision about co-location is affected by the type of workload hosted within VMs, CPU capacity, memory availability, and communication pattern of VMs. Degree of SLA violation is increased if infeasible VMs are co-located. Therefore, it is must to decide which VMs should be co-located. Parameters, such as, application profiling and statistical analysis helps in identifying the most suitable VMs.  
  \item Nowadays, power consumption within cloud data centers is a big challenge due to high power consumption by DC equipment because of hosting and deploying high resource demanding applications within data centers. Server consolidation is a mechanism that packs maximum possible VMs on a single server so that rest of the servers can be switched off to minimize power consumption budget. Moreover, applying dynamic voltage frequency scaling (DVFS) also helps to minimize power consumption budget. However, decisions about where to place services while keeping in mind customer's location and needs is a big  challenge that needs significant attention to surge data center performance.
  \item Lightweight VM migration design will help to minimize the resource usage of VM migration process as lightweight design uses low system resources. Incorporating  lightweight design feature in current VM migration schemes will help to accelerate DC performance. Moreover, proposing three dimensional queuing modeling based approach can be proposed to effectively highlight the objective and constraints of VM migration technology. 
\end{itemize}

\section{Queuing Based VM Migration Model}\label{sec:model}
\begin{figure*}
\centering
\includegraphics[width=14cm]{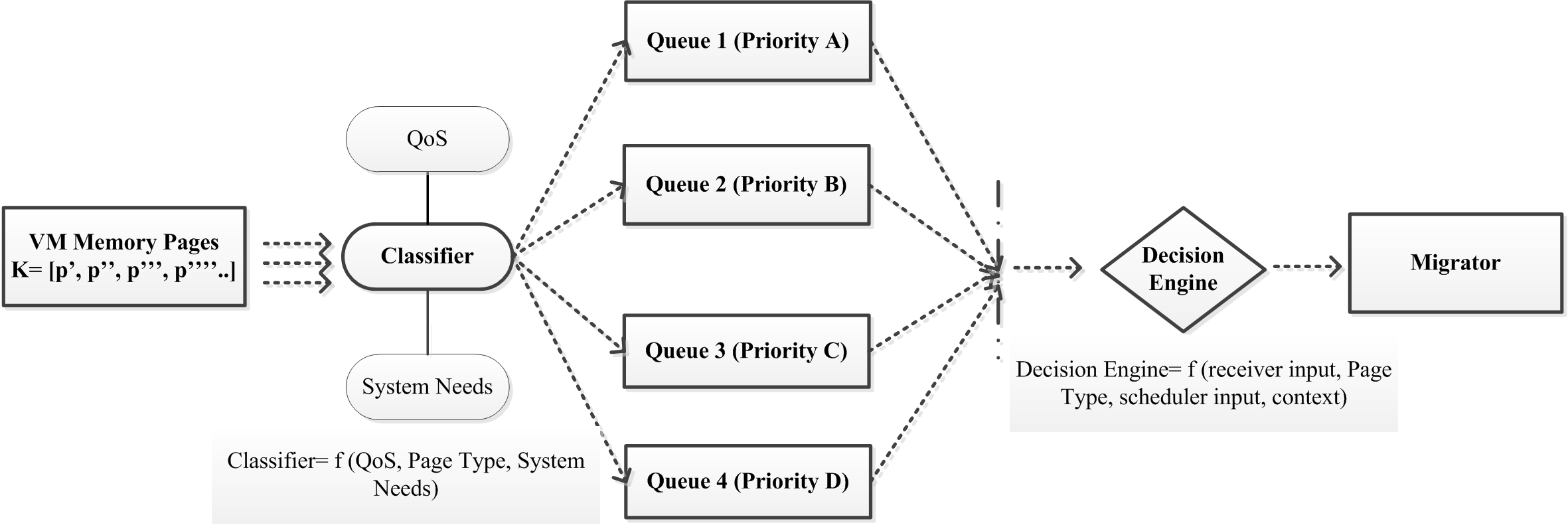}
\caption {Priority Queue based Migration Model}
\label{fig:queue}
\end{figure*}

This section discusses a queue based migration model that dynamically updates the VM memory pages queues and decides migration based on most needed pages at receiver end. Figure~\ref{fig:queue} shows the abstract presentation of proposed model.

In figure~\ref{fig:queue}, VM memory pages are classified into several classes based on the type of memory pages (e.g., pages needed to run VM or workload specific pages). Each class of memory pages is assigned some priority level that represents that whether those pages are required at the receiver end immediately or not. For instance, the memory pages which are immediately needed at receiver end are pushed in the high priority queue so that they can be transferred to the receiver end at the earliest. In the said figure, top queue (queue 1) represent the highest priority queue which is emptied as soon as possible (scheduler schedules their transferring again and again). Decision Engine is responsible to adeptly decide transferring of memory pages such that the element of fairness is not compromised. It selects a particular queue and trigger migrator module to migrate all the pages to the receiver until not stopped by decision engine again. The priority queues are formed adaptively as per system, application, and service demands. The round robin based schedule assigns heterogeneous time-slots to each queue while analyzing (by decision engine) the current needs and requirements of the application and system.  

\section{Conclusion and Future Works}\label{sec:conc}
Virtualization technology assists to achieve computing-as-a-service vision of cloud-based solutions. Virtual machine process helps to achieve various management goals, such as, load balancing, fault tolerance, and green cloud computing. It transfers system state from one server to another to offer uninterrupted services. However, VM migration is not free and consumes a significant amount of sender and receiver resources to carry out migration process successfully. This article has comprehensively discussed various basic elements of VM migration with special emphasis on VM migration applications and open issues. In the future we planned to implement queuing based VM migration model.  

\bibliographystyle{IEEEtran}
\bibliography{dc}
%

\end{document}